\documentclass[10pt,twocolumn,preprintnumbers,superscriptaddress,showpacs]{revtex4-1}
\usepackage{graphicx,amsmath,amssymb}

\usepackage[
       colorlinks=true,
      filecolor=black,
       anchorcolor=blue,
      linkcolor=blue,
      citecolor=red,
      urlcolor=blue,
       linktocpage=true,
        plainpages=false,
        breaklinks=true,
            pdfstartview=FitH
           ]{hyperref}
\usepackage{verbatim}
\usepackage{yfonts}
\usepackage{color}
\usepackage{ifsym}
\usepackage{lipsum}
\usepackage{hyperref}
\usepackage{titlesec}  

\def\xi{z}
\def\w{w}
\def\R{L}
\def\E{ {EPR}}
\def\i{{\mathrm{i}}}

\def\d{\mathrm{d}}
\def\tx{\tilde{x}}
\def\ty{\tilde{y}}
\def\tz{\tilde{z}}
\def\tg{\tilde{g}}
\def\tu{\tilde{r}}
\def\tw{\tilde{w}}
\def\tt{\tilde{\tau}}
\def\td{\tilde{q}}
\def\dt{{\cdot}}
\def\cO{{\cal{F}}}

\def\nn{\nonumber}
\def\no{n_\omega}

\def\i{{ \text{i}}}

\def\mR{\mathcal{R}}
\def\mA{\mathcal{A}}

\newcommand{\<}{\langle\,}
 \renewcommand{\>}{\rangle}
 \def\label{\label}

\renewcommand{\(}{\left(}
\renewcommand{\)}{\right)}

\begin{document}

\title{Bell Inequality in the Holographic EPR Pair}
\author{Jiunn-Wei Chen}\email{jwc@phys.ntu.edu.tw} 
\affiliation{Department of Physics, Center for Theoretical Sciences, and Leung Center for Cosmology and Particle Astrophysics, National Taiwan University, Taipei 10617, Taiwan}
\affiliation{Center for Theoretical Physics, Massachusetts Institute of Technology, Cambridge, MA 02139, USA}

\author{Sichun Sun}\email{Sun.Sichun@roma1.infn.it}
\affiliation{Department of Physics, Center for Theoretical Sciences, and Leung Center for Cosmology and Particle Astrophysics, National Taiwan University, Taipei 10617, Taiwan}
\affiliation{Department of Physics, Sapienza University of Rome, Rome I-00185, Italy}

\author{Yun-Long Zhang}\email{yun-long.zhang@yukawa.kyoto-u.ac.jp}
\affiliation{Center for Gravitational Physics, Yukawa Institute for Theoretical Physics, Kyoto University, Kyoto 606-8502, Japan}
\affiliation{Asia Pacific Center for Theoretical Physics, Pohang 790-784, Korea}
\affiliation{Center for Quantum Spacetime, Sogang University, Seoul 121-742, Korea}

\allowdisplaybreaks

\date{February 25, 2019}
\preprint{MIT-CTP/4869}

\begin{abstract}


We study the Bell inequality in a holographic model of the casually disconnected Einstein-Podolsky-Rosen (EPR) pair. The Clauser-Horne-Shimony-Holt(CHSH) form of Bell inequality is constructed using holographic Schwinger-Keldysh (SK) correlators. We show that the manifestation of quantum correlation in Bell inequality can be holographically reproduced from the classical fluctuations of dual accelerating string in the bulk gravity. The violation of this holographic Bell inequality supports the essential quantum property of this holographic model of an EPR pair.

\end{abstract}

\pacs{11.25.Tq,
03.65.Ud
\qquad arXiv: \href{https://arxiv.org/abs/1612.09513}{1612.09513}\quad
\quad DOI:  \href{https://doi.org/10.1016/j.physletb.2019.02.012}{10.1016/j.physletb.2019.02.012} }

\maketitle


 
\section{Introduction}
 
Bell inequality plays an important role in quantum physics \cite{Bell:1964kc}. Because correlations in local classical theories are bounded by the Bell inequality, which can be violated by the presence of the non-local entanglement in quantum mechanics.
The violation of  Bell inequality in the entangled Einstein-Podolsky-Rosen (EPR) pair \cite{Einstein:1935rr} indicates that two particles have an ``instant interaction'', 
in contrast to theories of hidden variables that preserve strict locality~\cite{Bell:1964kc, Einstein:1935rr, CHSH, Cirelson, Hartle:1992as}. The Bell inequality is a very general test in quantum physics. In addition to the Bell's tests in laboratories, there are also discussions of detections in cosmological scales, to confirm that whether the inflated primordial fluctuations are quantum mechanical in nature \cite{Maldacena:2015bha,Choudhury:2016cso,Chen:2017cgw}. 

In general relativity, two distant black holes can be connected by the wormhole or Einstein-Rosen (ER) bridge \cite{Einstein:1935tc}. Recently, Maldacena and Susskind proposed the ER=EPR conjecture ~\cite{Maldacena:2013xja,Susskind:2016jjb}, which suggested that the ER bridge can be interpreted as maximally entangled states of two black holes that form a complex EPR pair. This conjecture was proposed to resolve the  Almheiri-Marolf-Polchinski-Sully (AMPS) paradox without resorting to a firewall \cite{Almheiri:2012rt}. 
Interestingly, the ER=EPR conjecture implies that entanglement, which is thought to be a quantum mechanical effect of the EPR pair, can be captured by a complex ER bridge in gravitational theory.

%


Inspired by the original ER=EPR conjecture, a concrete holographic model of the EPR pair was proposed in Ref \cite{Jensen:2013ora}, based on the Anti-de Sitter/Conformal Field Theory(AdS/CFT) correspondence \cite{Maldacena:1997re} and one exact solution of the accelerating string in AdS$_5$ spacetime \cite{Xiao:2008nr}. 
In the holographic model, the EPR pair with two accelerating quasiparticles on the boundary
 is proposed to dual to an uniformly-accelerating open string in AdS$_5$ spacetime and the endpoints of the string are attached to the boundary.
Therefore, it is a holographic realization of the EPR pair, 
with an effective wormhole induced on the worldsheet in the AdS bulk. 
One should notice that the holographic setting is different from the original ER=EPR conjecture in which both ER and EPR live in the same spacetime dimensions.

%
In this paper, 
we focus on the Bell inequality in the holographic EPR model. 
We will show that the Bell inequality violated by the EPR pair living on the boundary 
can be captured by the perturbations of the dual accelerating string  with an effective wormhole on the worldsheet living in the AdS bulk spacetime.
To test the entanglement of the EPR pair, Bell inequality is a natural choice as it provides a sharp test of entanglement.
Since in the holographic model of an EPR model \cite{Jensen:2013ora}, it shows that the two quasi-particles are indeed entangled with each other, which indicates the violation of the Bell inequality. It is reasonable that one quantity in the holographic model can be indentified as the correlator in the Bell inequality, and our proposal is that the holographic Schwinger-Keldysh (SK) correlator can just play the role.  Thus, the violation of holographic Bell inequality supports the quantum entanglement property of this holographic model of an EPR pair.


In the following section \ref{Bell}, we firstly review the Bell inequality in quantum mechanics as well as in the field theory. In section \ref{EPR}, we summarize the holographic model of the EPR pair.
Our main result is in section \ref{hBell}, where we construct the Bell inequality based on this holographic model. We discuss and conclude the results in section \ref{Con}.
In the appendix \ref{App1}, we give more details on the derivation of the holographic Schwinger-Keldysh correlators.

\section{Bell Inequality in Quantum Theories}\label{Bell}
In this section, we will review the essence of Bell inequality, which is captured in the Clauser-Horne-Shimony-Holt (CHSH) correlation parametrizations \cite{CHSH}. 
The entangled states made of a pair of spin $1/2$ particles are detected by two observers, Alice and Bob, respectively. While the generalization to particles of higher spin is straightforward.
The operators correspond to measuring the spin along various axes with outcomes of eigenvalues $\pm 1$.  
Performing the operations $A$ and $A'$  on the first particle at Alice's location,
and operations $B$ and $B'$ on the second particle at Bob's location. 
With the Pauli matrices $\vec{\sigma}=({\sigma_x, \sigma_y, \sigma_z})$, and the unit vector $\vec{n}=(n_x,n_y,n_z)$ to indicate the spatial direction of the measurement, we have the following operators
 \begin{align}
A_s&=   \vec{n}_A\dt \vec{\sigma} ,~
A'_s=   \vec{n}_{A'}\dt \vec{\sigma} ,\\
B_s&=  \vec{n}_B\dt \vec{\sigma},~  
B'_s=  \vec{n}_{B'} \dt \vec{\sigma}.
\end{align} 
Then the {CHSH correlation formulation} is introduced as 
\begin{align} \label{eCHSH}
\langle  C_s \rangle =  \langle A_s B_s \rangle   + \langle A_s B'_s \rangle   + \langle A'_s B_s \rangle  - \langle  A'_s B'_s \rangle,
\end{align} 
which is a linear combination of crossed expectation values of the measurements. 
 
In a local theory with hidden variables, the formula is bounded by the Bell inequality $|\langle C_s \rangle | \leq 2$.  
While in quantum mechanics, this inequality can be violated, with a higher bound $|\langle C_s \rangle |\leq 2\sqrt{2}$ \cite{Cirelson} (see also \cite{Maldacena:2015bha} for the cosmological case). 
For example, if we choose the entanglement state of a spin singlet
\begin{align}\label{eEPR}
|\psi_s \rangle &= \frac{1}{\sqrt{2}}\big( |\!\uparrow\rangle \otimes |\!\downarrow\rangle - |\!\downarrow\rangle \otimes |\!\uparrow\rangle \big),
\end{align}
and take the measurements along the $(x,y)$ plane, i.e. $n_A=(\cos\theta_A, \sin\theta_A, 0)$ etc.,
it is straightforward to show
\begin{align} \label{Cdef}
G_{AB}^s \equiv \langle\psi_s | A_s B_s |\psi_s \rangle&= - \cos \theta_{A B}.
\end{align}
Here $\cos\theta_{A B} = \vec{n}_A \dt \vec{n}_B$ depends on the relative angle of the measurements.
And from \eqref{eCHSH} we have
\begin{align}  
  \langle\psi_s|C_s |\psi_s \rangle& ={-}\cos \theta_{A B}{-}\cos \theta_{A B'}{-}\cos \theta_{A' B}{+}\cos \theta_{A' B'}.
\end{align}

In particular, if we fix the direction of $A$ and $A'$, as well as the angle between $B$ and $B'$ as $\pi/2$,
\begin{align} \label{Achoices} 
\theta_A=0,\quad\theta_{A'}=\frac{\pi}{2},\quad \theta_{B'}=\theta_B-\frac{\pi}{2},
\end{align}
then we have the relation depended on the direction of $B,B'$,
\begin{align}
 \langle\psi_s |C_s |\psi_s\rangle =- 2 \sqrt{2}  \cos \big( \theta_B - \frac{\pi}{4}\big).
\end{align} 
For $0{<}\theta_B{<}\pi/2$,  the Bell inequality $ |\langle C_s \rangle | \leq 2$ can be violated,
and we reach the maximal violation at $\theta_B = \pi/4$, with an extra factor of $\sqrt{2}$.

{\it Bell's Test in Field Theory}--- 
Before we move to the holographic model, it is instructive to discuss the Bell's test in the context of the quantum field theory.
The test can be described by a process with the transition amplitude
\begin{align}
 \mathcal{A}=\< A^\uparrow B^\uparrow |T[P_A P_B]|\psi_s\> ,
 \end{align}
where a spin singlet state of Eq.(\ref{eEPR}) is measured by the projection operators 
\begin{align}
P_A=\frac{1+\vec{n}_A{\cdot}\vec{\sigma}}{2},\quad 
P_B=\frac{1+\vec{n}_B{\cdot}\vec{\sigma}}{2},
\end{align}
to the final state $|A^{\uparrow}B^{\uparrow} \>$ which is both spin up in the  $\vec{n}_A$ and $\vec{n}_B$ directions. Note that the operators are time-ordered since one can either measure $P_A$ or $P_B$ first. Squaring the amplitude to get the probability 
\begin{align}\label{timeordered}
\mathcal{P} \propto | \mathcal{A}|^2
&=\sum_X \< \psi_s|T[P_A P_B]^{\dagger}| X\>\< X|T[P_A P_B]|\psi_s\> \nn\\
&=\< \psi_s|T[P_A P_B]|\psi_s\>,
 \end{align}
 where only $|A^{\uparrow}B^{\uparrow} \>$ contributes in the complete set of state $| X\>$ in the first line and we have used the commutativity of $P_A$ and $P_B$, as well as $P_A^2=P_A$ and $P_B^2=P_B$ for projectors. 
 
The above discussion shows that the Bell's test is measuring a time-ordered Green's function which does not have to vanish when $P_A$ and $P_B$ are outside of each other's light cone. A familiar example of this is Feynman propagator which is also a time-ordered Green's function. Typically, this dramatic property of the Green's function does not show up in physical observable. It is very interesting that Bell's test relates those Green's functions to observables through Eq.(\ref{timeordered}). 
In the following holographic EPR model, we will study the  contour time-ordered Schwinger-Keldysh correlators of the EPR pair holographically.

\section{Holographic Model of an EPR Pair}\label{EPR}
In this section, we summarize the holographic model of an EPR pair in Ref. \cite{Jensen:2013ora}.
%
The holographic model proposed that  an entangled color singlet quark anti-quark ($q$-$\bar{q}$) pair in $\cal{N}$=4 supersymmetric Yang-Mills theory (SYM)
can be holographically described by an open string with both of its endpoints attached to the boundary of AdS$_5$. 
The string connecting the pair is dual to the color flux tube between the two quarks, with a $1/r$ Coulomb potential as required by the scale invariance of boundary theory.  Note that there is no confinement in this theory, therefore the pair can separate arbitrarily far away from each other. 

It was shown in \cite{Sonner:2013mba} that the holographic EPR model  corresponds to the Lorentzian continuation of the holographic Schwinger model. Since the particles are produced from the vacuum, they are necessarily entangled with one another, no matter what the actual nature of the particles,
may they be 
quarks and antiquarks and even charged $W_\pm$ bosons.  Various studies of related models can also be found in \cite{Jensen:2014bpa,Jensen:2014lua,Chernicoff:2013iga,Hirayama:2010xi,Caceres:2010rm}.

There are numerical solutions of the string shapes with different boundary behaviors~\cite{Herzog:2006gh,Chernicoff:2008sa,Chesler:2008wd}. But it is more convenient for us to work with the analytic solution for an accelerating string treated in the probe limit such that the back reaction to the AdS geometry is neglected~\cite{Xiao:2008nr}. 
In this analytic solution, the open string is also accelerated on the Poinc\'are patch of the AdS$_5$ spacetime
\begin{align}\label{metric}
{\d}s^2=\frac{\R^2}{\w^2}\big[-{\d}t^2+{\d}\w^2 +({\d}x^2+{\d}y^2+{\d}z^2)\big],
\end{align}
with the AdS radius $\R$ and extra dimension with coordinate $\w$. 
The string solution in the AdS$_5$ bulk is governed by the semicircle
\begin{align}\label{solution}
\xi^2=t^2+b^2-\w^2. 
\end{align}
For infinitely heavy test quarks, the quark and anti-quark live on the AdS boundary $\w=0$.
They are accelerating along the $\pm \xi$ direction, respectively, with the trajectories $\xi=\pm\sqrt{t^2+b^2}$.
Therefore, the two entangled particles are out of causal contact with each other the whole time. 
For the dynamical heavy quarks with the finite mass $m$, the string needs to end on a flavor probe brane \cite{Karch:2002sh,Karch:2013gsa}  at a small but finite $z_m=\frac{{T_s}\R^2}{m}<b$,
where $T_s$ is the tension of the string. A constant electric field $E=m/b$ on the flavor brane is responsible for accelerating the quasiparticles \cite{Jensen:2013ora}.

{\it String fluctuations.---}
To consider the string fluctuations, we transform the solution to the two co-moving frames of the accelerating particles with coordinates $(\tt, \tu, x,y,\tz)$ via
\begin{align} \label{transformation}
|\xi| &=  b \sqrt{1- \tu} {e^{\tz}} \cosh \tt, \quad 
t  =b \sqrt{1- \tu}{e^{\tz}} \sinh \tt,  \nn \\
\w &=b  \sqrt{ \tu} {e^{\tz}},\quad x=b\, \tx, \quad y=b\, \ty .
\end{align}
These two frames, which cover the regions $\xi\geq{0}$ and $\xi\leq{0}$ separately, are accelerating frames with a constant acceleration $a=1/b$ along opposite directions of $\xi$. 
And \eqref{transformation} only maps the upper part of the string ($0<w<b$) into the proper frames of the accelerating particles with $0<\tu< 1$. 
Plugging this transformation  \eqref{transformation} in the string solution \eqref{solution}, 
one finds the string configuration becomes $\tz=0$ for both frames of the quark and anti-quark.  Under this transformation  \eqref{transformation}, the metric \eqref{metric} becomes
\begin{align}  \label{newmetric} 
{\d}s^{2}{=}{\frac{\R^{2}}{ {\tu}}}\Big[{-}f({\tu}){\d}\tt^{2}{+}\frac{1}{4{\tu} }\frac{\d{\tu}^{2}}{f({\tu})}{+}{e^{-2\tz}} \( {\d}\tx^{2}{+}{\d}\ty^{2}\){+}{\d}{\tz}^{2} 
\Big],
\end{align}
where $f(\tu)=1-{\tu}$. 
There are the bifurcate Killing horizons at  $\tu=1$ associated with both frames of the accelerating quark and anti-quark.
Furthermore, with respect to the time $\tau {=} b \tt $,  
the Unruh temperature ${T_a}=\frac{1}{2\pi b}=\frac{a}{2\pi}$ \cite{Xiao:2008nr,Caceres:2010rm}.
We have set the reduced Planck constant $\hbar$ and Boltzmann constant $k_B$ to be unit. 
The regular thermal state with this temperature is similar to the Hartle-Hawking state in this spacetime with the Unruh vacuum,
and more details are addressed in the appendix \ref{App1}.

As shown in Fig. \ref{fig:hbell}, the two horizons are connected by part of the string which can be seen as an effective ER bridge on the string worldsheet.  It is suggested to be a holographic realization of the EPR pair in \cite{Jensen:2013ora},
although the dimensions are different from the original ER=EPR conjecture~\cite{Maldacena:2013xja,Susskind:2016jjb}. In this holographic model, EPR lives at the boundary with $3+1$ dimension while the ``effective ER'' lives on the worldsheet in bulk with $4+1$ dimension.

\section{Constructing Holographic Bell Inequalityl} 
\label{hBell}
In this section, we construct Bell Inequality in the holographic model of an EPR pair in \cite{Jensen:2013ora}.
The spin measurement of the particles can usually be carried out by the Stern-Gerlach type experiment which applied a magnetic field gradience to generate a force that acts on the spins of the particles. We expect similar measurement can be carried out in the holographic EPR model, which introduces fluctuations to the world lines of quasiparticles which set the boundary conditions for the {worldsheet} of the string fluctuations.
Let $\tx^a=(\tt ,{\tu})$ be the new worldsheet coordinates in the comoving frame, then the string fluctuation is described $X^{\mu }(\tt ,{\tu})=\big( \tt,{\tu}, \td_i (\tt ,{\tu})\big)$,  with  $i=(\tx,\ty,\tz)$.
When the fluctuations ${\td_i }\ll 1$, the Nambu-Goto action of string with tension ${T_s}$ becomes%
\begin{align}\label{action}
S_{NG} \!\simeq\!  -{T_s}\R^{2}\int \frac{d{\tt}d{\tu}}{2\tu^{ 3/2}}\left\{ 1+
\Big[2{\tu}f(\tu) {\td_{i}}^{\prime}{\td_{j}}^{\prime}-\frac{1}{2f(\tu)} \dot{\td}_{i}  \dot{\td}_{j}\Big] h^{ij}\right\} ,
\end{align}
where  ${\td_i}^{\prime}\equiv \frac{\partial {\td_i} }{\partial \tu}$, $\dot{\td}_i \equiv \frac{\partial {\td_i} }{\partial \tt}$ and $h^{ij}=\text{diag}[1,e^{2\tz},e^{2\tz}]$. 
The classical equations of motion for the fluctuations on the string are 
 \begin{align}\label{EOM}
\partial _{{\tu}}\Big( \frac{2 f {\td_i }^{\prime }}{{\tu}^{1/2}}\Big)
-\partial _{\tt}\Big( \frac{\dot{\td}_i}{2f{\tu}^{ 3/2}}\Big) =0.
 \end{align}
Focus on the transverse fluctuations along $i=\tx,\ty$,
 \begin{align}
{\td_i }({\tt},{\tu}) =\int \frac{d\omega  }{2\pi }e^{ -i\omega  {\tt }} {\td_i }(\omega) Y_{\omega  }( {\tu}) ,
\end{align}%
where $\td_i(\omega) $ is the Fourier transform of
fluctuation on the boundary directions and $Y_{\omega  }( {\tu})$ is the function of the radio direction $\tu$ and frequency $\omega$  . 
%

\begin{figure}[h]
\begin{center}
\includegraphics[scale=0.21]{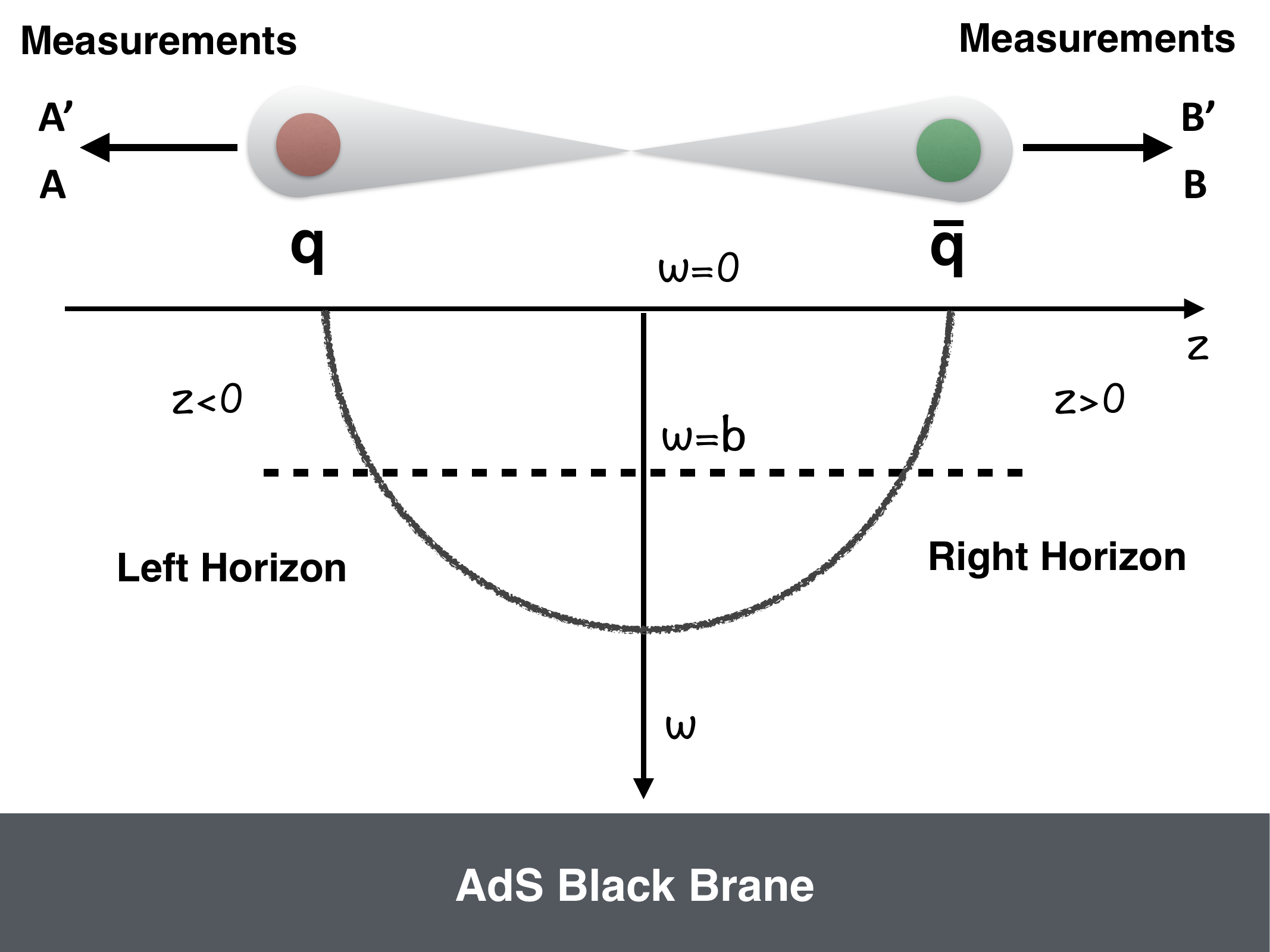}\\
\caption{Schematic diagram for the test quark($q$)-antiquark($\bar{q}$) EPR pair in holographic model at a fixed time $t>0$.
From \eqref{transformation}, the trajectory of left(right) {worldsheet} horizon is on $\w=b$, 
which depicts the intersect of the string with the world volume horizon seen by $q$($\bar{q}$) in its co-moving frame. The Bell inequality test is performed by the measurements at Alice and Bob's locations. The AdS {black brane} is present only in the finite temperature case, which we will briefly discuss in the last section. }\label{fig:hbell}
\end{center}
\end{figure}

The retarded Green's function of the particle under effective random force $\cO^i(\tau)$ can be defined as 
${\i}G_{\mathcal{R}}^{ij}(\tau)=\theta(\tau)\langle [\cO^i(\tau),\cO^j(0)] \rangle$ \cite{Xiao:2008nr}.
 In the AdS/CFT correspondence, $\cO^i(\tau)$ is the operator conjugate to the  fluctuations $ q_i(\tau)$ of the fundamental string,
where the dimensionful quantities are $\tau=b\tt, q_i=b\td_i$. 
In the low frequency limit $\omega\to 0$, it can be obtained analytically as
\begin{align}
G_{\mathcal{R}}^{ij}(\omega) &=-\frac{2{T_s}\R^{2} }{b^{2} \tu^{1/2}}
f( {\tu}) Y_{-\omega }( {\tu}) \partial _{{\tu}}Y_{\omega}( {\tu}) \delta^{ij}\big\vert _{{\tu}\rightarrow 0} \nn \\
&= -\frac{a^2 \sqrt{\lambda}}{2\pi  } \i\omega \delta^{ij}+O(\omega^2),
\end{align}%
and we have used the fact that ${T_s}\R^2=\frac{\sqrt{\lambda}}{2\pi}$.

What we need for the Bell's test is the contour time-ordered Schwinger-Keldysh (SK) Green's function,
\begin{align}\label{FAFB}
\i G_{AB}^{ij}(\tau, x)=\langle \cO_A^i(\tau, x) \cO_B^j (0) \rangle,
\end{align}
where $\cO_A^i$ and $\cO_B^j$ are separately defined on the causally disconnected left and right wedges of the Penrose diagram, corresponding to the boundaries of different patches of the AdS space.  
With these conjectures, in \cite{Herzog:2002pc}, it was shown that in the  semi-classical limit of the AdS/CFT correspondence, the two-point correlators can be calculated from the variations
of the generating function of the boundary quantum field theory,
which is identified with the total on-shell action of the dual field in the bulk gravity.
In our case, the bulk fields are the string fluctuations, 
which are governed by the equations of motion from the Nambu-Goto action in the bulk gravity,
with the corresponding  boundary conditions $\td_i(\tt,\tu)|_{\tu \to 0}{=}{\td}^I_i(\tt)$. 
Thus, through the AdS/CFT correspondence in semi-classical limit (or large-N, large 't Hooft coupling limit of the dual field theory), the generating function of the holographic EPR pair could be assumed to be 
\begin{align} \label{semiclassical}
Z_{\E} 
\equiv \langle e^{\frac{\i}{\hbar}S_{\E} }\rangle 
\overset{AdS/CFT}{\simeq}  e^{\frac{\i}{\hbar} S_{NG}[ {\td}^I_i, {\td}^J_j]} .\vspace{-5pt}
\end{align}

After considering the holographic advanced and retarded Green's functions in \eqref{GmR}  \eqref{GmA}  in Appendix \ref{App1},
taking functional derivatives of $Z_{\E}$ with respect to ${\td}^I_i $ and ${\td}^J_j $ will yield  the Schwinger-Keldysh(SK) correlators in the holographic EPR pair 
\begin{align} \label{FiFj}
& \i G_{IJ}^{ij}  \equiv \frac{\hbar^2}{\i^2} \frac{\delta^2\ln Z_{\E}}{\delta({\td}^I_i) \delta ({\td}^J_j) }
\simeq  \frac{\delta^2 S_{NG}[ {\td}^I_i, {\td}^J_j]}{\delta({\td}^I_i) \delta ({\td}^J_j) }, 
\end{align}
where the semiclassical limit in \eqref{semiclassical} has been used. 
 
This  off-diagonal SK correlator for the EPR pair is found to be related to the holographic retarded Green's function
\begin{align}\label{GABomega}
& \i G_{AB}^{ij} ( \omega ) =\frac{-2e^{-\omega/(2{T_a})}}{1-e^{-\omega/{T_a}}} \textrm{Im}G_{\mR}^{ij}\left( \omega \right),
\end{align}
which has been examined in the equation \eqref{SKcr}.
It is similar to what was found for the boundary field theory in Refs. \cite{Son:2007vk,Herzog:2002pc,Son:2002sd} 
but that is from the on-shell action of the bulk gravity.  
From the boundary theory's viewpoint, the operators $\cO_A^i, \cO_B^i$ in \eqref{FAFB} conjugated to the sources ${\td}^A_i, {\td}^B_i$ are quantized through path integral quantization.
However, in the semi-classical limit of the AdS/CFT correspondence for the string worldsheet in \eqref{semiclassical}, the final result of the path integral of this  quantum EPR theory is assumed to be the exponentiated classical on-shell Nambu-Goto action $S_{ NG } [{\td}^I_i, {\td}^J_j]$ in \eqref{NGb2}.
 From the bulk theory's viewpoint,  the string in the AdS bulk is in the probe limit and the string fluctuations are still governed by the classical equations of motion in \eqref{EOM},
which are obtained from the variation of the Nambu-Goto action in \eqref{action}.

For fluctuations coming from two causally separated quarks of an EPR pair along $x$ and $y$ directions,
and in the low frequency limit $\omega\to 0$,
\begin{align}\label{Green}
&\i G_{AB}^{xx} = \i G_{AB}^{yy} =  \frac{\sqrt{\lambda} a^3}{2\pi^2} , \quad  \i G_{AB}^{xy}=\i G_{AB}^{yx}=0, 
\end{align}
which indicates that the spatial correlator $G^{ij}_{AB} \propto \delta^{ij}$.
The $\sqrt{\lambda}$ factor is consistent with the observation that the entanglement entropy of the  entangled pair is of order $\sqrt{\lambda}$ \cite{Jensen:2013ora}. It is also interesting that this SK correlator does not vanish when the particles are separated at long distance. 
This is consistent with the non-local nature of entanglement. 
However, 
the SK correlator vanishes when the acceleration $a$ becomes zero, when the ER bridge on the worldsheet disappears. Although it is unclear that how the holographic Bell inequality is evaluated in this case, we may only approach the zero acceleration limit after we identify the correlation with the normalized operators as in the following Eq.(\ref{operator}) in which $a$ dependence cancels.

To study the holographic correlators, we normalize the force operators $\cO_I^i$ such that only the dependence on directions remains
%
\begin{align}
A_{\cO}=(\cos\theta_A\cO_A^x+\sin\theta_A\cO_A^y)/\langle \cO_A^x \cO_B^x \rangle^{1/2},\nn\\ 
B_{\cO}=(\cos\theta_B\cO_B^x+\sin\theta_B\cO_B^y)/\langle \cO_A^x \cO_B^x \rangle^{1/2}.\label{operator}
\end{align}
Similar to \eqref{Cdef}, the mixed measurements for correlators in {CHSH correlation formulation} become
\begin{align} \label{cAB}
\langle A_{\cO}B_{\cO} \rangle=
 \cos(\theta_{A}-\theta_{B})\equiv \cos\theta_{A B}.
\end{align}
Together with the similar normalization of the operators $A'_{\cO}$ and $B'_{\cO}$,
 the {CHSH correlation formulation} in \eqref{eCHSH} becomes
\begin{align} \label{CdefF}
\langle  C_{\cO} \rangle &=  \langle A_{\cO}B_{\cO}  \rangle   + \langle A_{\cO}B'_{\cO}  \rangle   + \langle A'_{\cO}B_{\cO} \rangle  - \langle  A'_{\cO}B'_{\cO} \rangle\nn\\
&=  \cos\theta_{A B}+ \cos\theta_{A B'\!}+ \cos\theta_{A'\!B}- \cos\theta_{A'\!B'\!}. 
\end{align} 
For example, when 
$\theta_{A B}=\theta_{A B'}=\theta_{A' B}=\pi/4$, and $\theta_{A'B'}=3\pi/4$,
we can reach the maximum value $2\sqrt{2}$.
It suggests that in this holographic EPR system, the CHSH formulation of Bell inequality $\langle  C_{\cO} \rangle\leq 2 $ can be violated, which supports the essential quantum property of the holographic EPR pair in Ref. \cite{Jensen:2013ora}.



\section{Conclusion and Discussions} \label{Con}

In this paper, using Bell inequality as a sharp test of entanglement, we study a holographic model with an 
EPR pair at the boundary theory and an effective wormhole on the string worldsheet in the bulk.
By revealing how Bell inequality is violated in the EPR pair through the holographic duality from the on-shell string fluctuations in higher dimension bulk gravity, our study supports the essential property of this holographic model of EPR pair in Ref. \cite{Jensen:2013ora}. We also explained that why a higher dimensional classical theory can reproduce the quantum property of the boundary theory. Because from the boundary theory's viewpoint, the operators conjugated to the fluctuations of the EPR pair are quantized through path integral quantization.
While in the semi-classical limit of AdS/CFT correspondence, the final result of the path integral of this quantum EPR theory is assumed to be the exponentiated classical on-shell Nambu-Goto action.
Thus, from the bulk theory's viewpoint, the string in the AdS bulk is in the probe limit and the string fluctuations are governed by the classical equations of motion, which are obtained from the variation of Nambu-Goto action.

In our derivation, we see that the bulk string fluctuations in the AdS gravity produce the holographic correlators. We assume that the holographic SK correlators play the role as that in the CHSH formulation of the  Bell inequality in a dual EPR pair on the boundary, which can be violated.  
Technically, this result relies on only two ingredients. The first one is that the observable in Bell's test is a time-ordered Greens function as shown in Eq.( \ref{timeordered}). And it is well known that the time-ordered Green's function does not have to vanish when the measurements $P_A$ and $P_B$ are outside of each other's light cone. Mathematically, this is because the behaviors of the SK correlators outside the two horizons need to be correlated, otherwise the solution is not smooth inside the horizons.
The second one is that the equation of motion of the classical string, Eq.(\ref{EOM}), has no coupling between $\td_x$ and $\td_y$ such that Eq.(\ref{Green}) follows. This can be obtained as long as the string does not experience a force to propagate the fluctuation in the $x$-direction to the $y$-direction which breaks parity in general. 
It seems once the two conditions above are satisfied, it does not matter whether there is an effective wormhole on the worldsheet in the bulk. 
Hence it is conceivable Bell inequality can still be violated in a holographic model where the EPR pair does not accelerate,  similar to how holographic entanglement entropy is computed in a static system \cite{Ryu:2006bv,Numasawa:2016emc}.

In this paper we are only work with the holographic model of an EPR pair. 
For future work, it is interesting to consider the back reaction by the measurements and see whether the effective wormhole on the worldsheet is broken due to the energy injected by measurements. This might provide an opportunity to study the ``wave function collapse" typically used to describe how measurements change the states.
Another interesting topic is the decoherence of the EPR pair in the environment. If the environmental effect can be described by thermal fluctuations, then we can add a black hole to the bulk of our model. When the distance between the particles in the EPR pair increases with time, the effective wormhole on the worldsheet also approaches the black brane horizon and then enters the horizon~\cite{Dominguez:2008vd}. We expect the string breaks after it enters the horizon which might shed light on the decoherence process in the boundary field theory. For the current holographic model of EPR pair we studied, the SK correlators will vanish in the limit of zero acceleration $a\to 0$ and there is no ``effective ER bridge'' on the worldsheet anymore. Although this does not necessarily contradict with the ER=EPR conjecture, since there may exists an ER bridge of the other sort for this static EPR pair. 

Our work is complementary to Refs.~\cite{Bao:2015nqa,Bao:2015nca,Remmen:2016wax} in which properties of quantum mechanical entanglement are given geometrical interpretations using holography. For example, the no-cloning theorem in the quantum mechanics is related to the no-go theorem for topology change in general relativity \cite{Bao:2015nqa}. The un-observability of entanglement for most of the observables is related to the undetectability of the presence or absence of a wormhole for most of the observables \cite{Bao:2015nca}. The entanglement conservation is related to the invariant of a certain area in general relativity \cite{Remmen:2016wax}. It will be interesting to further explore the connection of these results to our model on holographic Bell inequality.

It will also be interesting to see whether our work can be generalized to distinguish the quantum mechanical ``squashed''\cite{Squashed:Chr03} and the classical Shannon~ \cite{Shannon1948}  entanglement. While the violation of Bell inequality for an EPR pair implies the pair is entangled quantum mechanically,  it is non-trivial to generalize such a test from an EPR pair to a many-body system described by the boundary CFT. We envision this kind of theory would be able to subtract the classical (e.g. thermal) contribution to the Ryu-Takayanagi entanglement entropy formula  \cite{Ryu:2006bv,Numasawa:2016emc}.

Finally, the strongest version of ER=EPR suggests that each EPR pair is literally connected by an ER bridge \cite{Maldacena:2013xja}, not just dual to an ER bridge. So both ER and EPR coexist in the same unified theory of the same dimensions, in contrast to our holographic model where ER and EPR exist in two different theories of different dimensions. Moving the EPR pair of our model to the bulk (which might involve quantizing the string fluctuations) might provide a handle on studying this problem.

\vspace{5pt}

\appendix
\section{Holographic SK Correlator}  
\label{App1}
In this appendix, we give a detailed derivation of the relation between holographic Schwinger-Keldysh correlator
and retarded Green function in equation \eqref{GABomega}. 
 The derivation  follows Ref. \cite{Herzog:2002pc,CasalderreySolana:2007qw}.
Let $\tx^\mu=(\tt ,{\tu},\tx,\ty,\tz)$ be the dimensionless coordinates in the co-moving frame of the of the accelerating particles via \eqref{transformation}.
The square of line element of the AdS spacetime becomes ${\d}s^2=\tg_{\mu\nu}{\d} \tx^\mu {\d} \tx^\nu$, where 
\begin{align}  \label{newmetric1} 
\tg_{\mu\nu}= \frac{\R^{2}}{{\tu}}  \text{diag}\big[{-}f({\tu}), \frac{1}{4{\tu} f({\tu})}, {e^{-2\tz}}, {e^{-2\tz}} , 1 \big]  .
\end{align}
And $\tx^a=(\tt ,{\tu})$ are coordinates on the static string worldsheet.
Without loss of generality,
we can consider the string fluctuation as $X^{\mu }(\tt ,{\tu})=\big( \tt,{\tu}, \td_{i} (\tt ,{\tu})\big)$, which lead to an induced metric on the string worldsheet $g_{ab}{=}(\partial_a X^\mu)(\partial_b X^\nu) \tg_{\mu\nu}$.
The Nambu-Goto action of string with tension  ${T_s}$ is
$S_{NG}{=-}T_s{\int} \d{\tt}\d{\tu}\sqrt{-\det{g_{ab}}} $.
When the fluctuation ${\td_i }\ll 1$, the action becomes%
\begin{align}\label{NGaction}
S_{NG}{\simeq}{-}\frac{\sqrt{\lambda}}{2\pi} {\int}\frac{\d{\tt}\d{\tu}}{2\tu^{ 3/2}}\left\{1{+}
\Big[2{\tu}f(\tu) {\td_{i}}^{\prime}{\td_{j}}^{\prime}{-}\frac{1}{2f(\tu)} \dot{\td}_{i}  \dot{\td}_{j}\Big] h^{ij}\right\} ,
\end{align}
where  $ {\sqrt{\lambda}}\equiv {2\pi}{T_s}\R^{2}$,  ${\td_i}^{\prime}\equiv \frac{\partial {\td_i} }{\partial \tu}$, $\dot{\td}_i \equiv \frac{\partial {\td_i} }{\partial \tt}$ and $h^{ij}=\text{diag}[1,e^{2\tz},e^{2\tz}]$ have been used. 

The classicial equations of motion for the fluctuations $\td_{i}$ on the string are 
 \begin{align}
\partial _{{\tu}}\Big( \frac{2 f {\td_{i} }^{\prime }}{{\tu}^{1/2}}\Big)
-\partial _{\tt}\Big( \frac{\dot{\td}_{i}}{2f{\tu}^{ 3/2}}\Big) =0. \label{stringeom}
 \end{align}
Performing a Fourier transform, 
 \begin{align}\label{fourier}
{\td_{i}}({\tt}, {\tu}) =\int \frac{d\omega  }{2\pi }e^{ -i\omega  {\tt }} {\td_i }(\omega) Y_{\omega  }( {\tu}), 
\end{align}%
where $\td_i(\omega) $ is defined as the Fourier transform of
fluctuation on the boundary,
after choosing the normalization $\lim_{\tu\to 0}Y_{\omega  }( {\tu }) =1$. 
Then \eqref{stringeom} becomes
\begin{align}\!\!
Y''_{\omega}( {\tu})- \frac{f(\tu)-2\tu{f'(\tu)}}{2 \tu f(\tu)}Y'_{\omega}({\tu}) +\frac{\omega^2 Y_{\omega}( {\tu})}{2f(\tu)\tu^{3/2}}=0.
\end{align}%
Requiring the in-falling boundary condition at the horizon, this equation is solved by
\begin{align}
Y_{\omega}( {\tu})=(1-\tu)^{-\i\omega/2} F_{\omega}({\tu}).
\end{align}
The complex conjugate 
\begin{align}
Y^*_{\omega}( {\tu})=Y_{-\omega}( {\tu})=(1-\tu)^{+\i\omega/2} F_{-\omega}({\tu})
\end{align}
is the other solution with the outgoing boundary condition at the horizon.

\begin{figure}[h]
\includegraphics[scale=0.15]{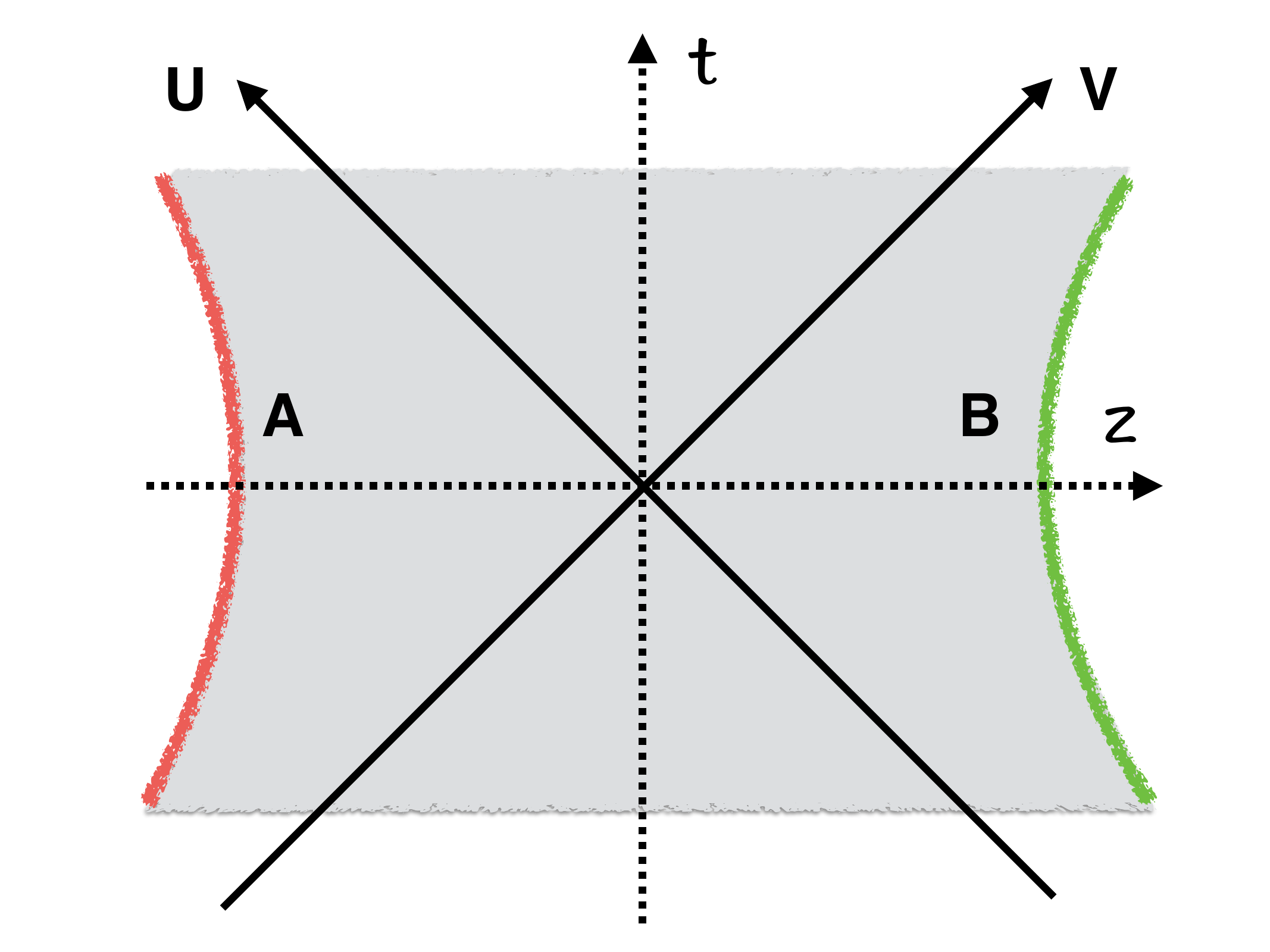} 
\caption{The Kruskal plane of the  metric \eqref{newmetric}  in terms of coordinates $U$ and $V$.
The string worldsheet covers  four quadrants, with the time-like boundary A(Red) at the left quadrant, and B(Green) at the right quadrant.}\label{fig:UV}
\end{figure}

We need to extend these solutions into the Kruskal plane of the  metric \eqref{newmetric1},
with new coordinates $U$ and $V$, which are initially defined in the right-quadrant $\{{U<0}, {V>0}\}$,  with
\begin{align}
{U}&=- e^{-2\tt} e^{-2\tu^*},\quad
{V} =+ e^{+2\tt} e^{-2\tu^*}.
\end{align}
And $\tu^*$ is placed outside the worldsheet horizon ${0\!<\!\tu\!<\! 1}$,
\begin{align}
\tu^* \equiv \int_0^{\sqrt{\tu}}\frac{ d\tw }{f(\tw^2)} =\frac{1}{2} \ln\frac{1- {\sqrt{\tu}}}{1+\sqrt{\tu}}, 
\end{align}
with ${f(\tw^2)} =1- \tw^2$.
The full extension of the  metric \eqref{newmetric1} and string worldsheet  in the Kruskal plane is shown in Fig. \ref{fig:UV}.

In the right-quadrant the two solutions near the horizon are 
\begin{align}\label{Bi}
\td^B_- &=e^{-i{\omega} {\tt}}{{Y}_{\omega}(\tu)} \sim  e^{-i(\omega/2) \ln(V)} ,\\
\label{Bo}
\td^B_+&=e^{-i{\omega} {\tt}}{Y^*_{\omega}(\tu)}  \sim e^{i({\omega}/{2})\ln(-{U})}.
\end{align}
And in the left quadrant $\{{U>0}, {V<0}\}$, 
\begin{align}
\td^A_- &=e^{-i{\omega} {\tt}}{{Y}_{\omega}(\tu)} \sim  e^{-i(\omega/2) \ln(-V)} \label{Ai},\\
\td^A_+ &=e^{-i{\omega} {\tt}}{Y^*_{\omega}(\tu)} \sim e^{i(\omega/2) \ln(U)} \label{Ao}.
\end{align}
Similar to the Herzog-Son's prescription \cite{Herzog:2002pc,Unruh:1976db}, 
two linear combinations are analytic over the full Kruskal plane,
\begin{align}
\td_+(\omega,\tu)&=\td^B_+ + e^{+\pi \omega/2} \td^A_+  ,\\
\td_-(\omega,\tu)&=\td^B_-+ e^{ -\pi \omega/2} \td^A_- ,
\end{align}
which can be used as two bases for the string fluctuations,
\begin{align}\label{basis}
\td_i(\tt,\tu)=\int \frac{d{\omega}}{2\pi} \left[a_i(\omega)\td_+(\omega,\tu)+ b_i(\omega)\td_-(\omega,\tu)\right].
\end{align}
The coefficients $a_i({\omega}), b_i({\omega})$ can be determined by the two  boundary values  $\td_i^A(\omega) $ and $\td_i^B(\omega)$ of the solutions,
\begin{align} \label{cAB}
a_i(\omega)&=\no\big[-\td_i^A({\omega}) +e^{\pi\omega/2} \td_i^B({\omega})  \big], \\
b_i(\omega)&=\no\big[ e^{\pi{\omega}}\td_i^A({\omega})-e^{\pi\omega/2}\td_i^B({\omega})\big],
\end{align}
with $\no={1}/{(e^{\pi\omega}-1)}$.
 
In the following, we calculate the boundary formula of the on-shell
 Nambu-Goto action in \eqref{NGaction}.
After considering the classical equations of motions \eqref{stringeom} and integrating out the $\tu$ direction, 
we obtain
\begin{align}\label{NGb1}
 S_{ NG } &=  \frac{\sqrt{\lambda} f( {\tu}) }{{\pi} b^{2} \tu^{1/2}}\big(\!\int_A -\int_B\! \big)\frac{d{\omega}}{2\pi}  
\Big[ {\td}_i(-\omega , {\tu}) \partial _{{\tu}}{\td}_j({\omega}, {\tu}) \delta^{ij}\Big]\big\vert _{{\tu}\rightarrow 0}.
\end{align}
In \eqref{basis}, we have obtained the analytic and regular solutions for the string fluctuations.
Putting them back into \eqref{NGb1}, the on-shell  Nambu-Goto action becomes
\begin{align}\label{NGb2}
S_{ NG }& [{\td}^I_i, {\td}^J_j] 
 = \!-\! \frac{1}{2} \! \int \! \frac{d \omega}{2\pi} 
\Big\{ \big[{\td}^A_i(-\omega){\td}^B_j(\omega) + {\td}^B_i(-\omega){\td}^A_j(\omega) \big] \nn\\
&    \times  \sqrt{\no(1+\no)}  \big[ G_{\mA}^{ij}(\omega) -G_{\mR}^{ij}(\omega) \big]  \nn\\
& +{\td}^A_i(-\omega){\td}^A_j(\omega) 
\big[(1+n)G_{\mR}^{ij}(\omega) -n G_{\mA}^{ij}(\omega)\big] \nn\\
& +{\td}^B_i(-\omega){\td}^B_j(\omega)
 \big[ n G_{\mR}^{ij}(\omega) -(1+n) G_{\mA}^{ij}(\omega)\big]  \Big\},\!
\end{align}%
where $I,J{=}A,B$ and
\begin{align}
G_{\mR}^{ij}(\omega) &=-\frac{\sqrt{\lambda} f( {\tu}) }{{\pi} b^{2} \tu^{1/2}}
Y_{-\omega }( {\tu}) \partial _{{\tu}}Y_{\omega}( {\tu}) \delta^{ij}\big\vert _{{\tu}\rightarrow 0}, 
\label{GmR} \\
G_{\mA}^{ij}(\omega) &=-\frac{\sqrt{\lambda} f( {\tu}) }{{\pi} b^{2} \tu^{1/2}}
Y_{\omega }( {\tu}) \partial _{{\tu}}Y_{-\omega}( {\tu}) \delta^{ij}\big\vert _{{\tu}\rightarrow 0}.\label{GmA}
\end{align}%
This can be compared with  \cite{Son:2002sd}, where $G_{\mR}^{ij}$ and $G_{\mA}^{ij}$ are conjectured to be the holographic retarded and advanced Green's functions.
From \eqref{NGb2} and  \eqref{FiFj}, the off-diagonal one is related to the retarded Green's function as \eqref{GABomega} in the main text,
\begin{align} \label{SKcr}
\!\! \i G_{AB}^{ij} (\omega) {\equiv} \frac{S_{ NG }  [{\td}^I_i, {\td}^J_j] }{\delta({\td}^A_i) \delta ({\td}^B_j) }
{=}\frac{ - 2  e^{-\omega/(2T_a)}}{1-e^{-\omega/T_a}} \textrm{Im}G_{\mR}^{ij}\left( \omega \right). 
\end{align} 
If restoring the physical units, the exponential index becomes $\exp[-{  \frac{\hbar\,  \omega }{k_B T_a } }]$,
and ${T_a} = \frac{\hbar \, a}{  2\pi k_B c }$.
Thus, the Planck constant does not appear in the correlators.

\quad  

{\bf Acknowledgements}. ---
We are grateful to A. Karch for many valuable comments and discussions. 
We thank helpful comments from D. Berenstein, F. L. Lin, and R. X. Miao, as well as
many insightful comments from the referees, which all help us to finalize this version.
J. W. Chen and S. Sun were supported by the MOST and NTU-CTS at Taiwan.
J.W. Chen was also partially supported by MIT MISTI program and the Kenda Foundation. S.\, Sun was partially supported by MIUR in Italy under Contract(No. PRIN 2015P5SBHT) and ERC Ideas Advanced Grant (No. 267985) \textquotedblleft DaMeSyFla";\, 
Y. L. Zhang was supported by CQUeST, APCTP(funded by the Ministry of Science, ICT and Future Planning(MSIP), Gyeongsangbuk-do and Pohang City) and Grant-in-Aid for JSPS international research fellow(18F18315).


\end{document}